 \documentclass[prl,twocolumn,showpacs,twocolumngrid,superbib]{revtex4}
 \newcommand{\commentoutB}[1]{}
 \newcommand{\commentoutA}[1]{#1}
\usepackage{graphicx}
\usepackage{amsmath}
\usepackage{bm}
\usepackage{alltt}
\usepackage{fancyhdr}

\def\Tr{{\rm Tr}}
\def\F{\mathcal{F}}
\def\D{\mathcal{D}}
\def\X{\mathcal{X}}
\def\E{\mathcal{E}}

\begin{document}

\title{Higher Order Response in ${\cal O}(N)$ by Perturbed Projection} 

\author{Val\'ery Weber}
\email{valery.weber@unifr.ch}
\affiliation{Department of Chemistry, University of Fribourg, 1700 Fribourg, Switzerland.}
\author{Anders M. N. Niklasson}
\author{Matt Challacombe}
\affiliation{Los Alamos National Laboratory, Theoretical Division, Los Alamos 87545, New Mexico, USA.}

\date{\today}

\begin{abstract}
Perturbed projection  for linear scaling solution of the coupled-perturbed 
self-consistent-field equations 
[Weber, Niklasson and  Challacombe, Phys.\ Rev.\ Lett. {\bf 92}, 193002 (2004)] 
is extended to the computation of higher order static response properties.
Although generally applicable, perturbed projection is developed here 
in the context of the self-consistent first and second electric hyperpolarizabilities 
of three dimensional water clusters at the Hartree-Fock level of theory. 
Non-orthogonal, density matrix analogues of Wigner's $2n+1$ rule are given up to fourth order.
Linear scaling and locality of the higher order response densities under perturbation 
by a global electric field are demonstrated.  
\end{abstract}



\maketitle

\footnotetext[1]{LA-UR-04-5219} 

\section{Introduction}
First principles electronic structure theory has traditionally been limited 
to the study of small systems with a limited number of nonequivalent atoms. 
Despite the tremendous increase in computational power of digital computers this 
has remained the case, until the advent of reduced complexity algorithms over the
last decade \cite{GGalli96,DBowler97,SGoedecker99,POrdejon00,VGogonea01,SWu02}. In the 
best case, these reduced complexity algorithms scale only linearly with system size, $N$, 
allowing simulation capabilities to keep pace with hardware improvements.
These linear scaling algorithms exploit the quantum locality (or nearsightedness) of 
non-metallic systems,  manifested in the approximate exponential decay of density matrix elements 
with atom-atom separation through the effective use of sparse matrix methods. For small systems,
linear scaling methods may be inefficient due to overhead.  However, for large, complex systems
these methods hold the promise of major impact across materials science, chemistry and biology. 

So far, a majority of work in linear scaling electronic structure theory 
has focused on methods and calculations involving the ground state, with little 
attention devoted to the problem of response properties.  The calculation of static
response within  Hartree-Fock or Density Functional Theory may be  obtained through solution 
of the Coupled-Perturbed Self-Consistent-Field (CPSCF) equations, which yield properties 
such as the electric polarizability and hyperpolarizability \cite{HSekino86,SKarna91}, 
the Born-effective charge, the nuclear magnetic shielding tensor \cite{KWolinski90}, 
indirect spin-spin coupling
constant \cite{CPennington91,OMalkin96}, geometric derivatives (i.e.~higher order
analytic force constants) \cite{RAmos89} and polarizability derivatives such as the Raman 
intensity \cite{MLazzeri03,OQuinet01}, to name but a few.

Conventional approaches to solution of the CPSCF equations 
\cite{HSekino86,SKarna91,JPople79} are based on perturbation of the wave 
function, requiring an $N^3$-scaling eigensolve which may need to be followed by an ${\cal O}(N^5)$ 
transformation of two-electron integrals, depending on the method. 
In addition to the formal scaling of these conventional methods, they do not admit exploitation 
of quantum locality through the effective use of sparse matrix algebra.  
More recently, schemes with the potential for reduced complexity have been been put forward.
Ochsenfeld and Head-Gordon proposed a scheme based on the Li-Nunes-Vanderbilt 
density matrix minimization \cite{Ochsenfeld97}.  Later, Larsen {\em et al.} \cite{HLarsen01a} 
proposed iterative solution of the CPSCF equations involving equations derived from 
unitary operations and approximations to the matrix exponential.    In both of these approaches, 
a linear system of equations containing commutation relations is obtained, which {\em implicitly}
determines the response function.  However, the method of solution for these equations
is not discussed, and computational results are not presented.  Recently though, 
with an apparent reformulation of Ref.~[\onlinecite{Ochsenfeld97}],
Ochsenfeld, Kussmann and Koziol~\cite{COchsenfeld04} have achieved near linear scaling computation
of NMR chemical shifts for one-dimensional alcanes at the GIAO-HF/6-31G* level of theory,
but likewise provide no details on their method of solution.  These implicit commutation relations are Sylvester-like
and may be solved with a number of approaches \cite{JBrandts01}, the particulars of which are of interest. 

In contrast, Perturbed Projection \cite{VWeber04} is a recently developed alternative 
for $N$-scaling solution of the CPSCF equations that is simple and explicit.  Based on a recently developed density 
matrix perturbation theory \cite{ANiklasson04}, Perturbed Projection exploits the explicit 
relationship between the density matrix $\mathcal{D}$ and the effective Hamiltonian or 
Fockian $\mathcal{F}$ via spectral projection; $\mathcal{D}=\theta(\tilde{\mu}I-\mathcal{F})$, 
wherein $\theta$ is the Heaviside step function (spectral projector)  and the chemical potential 
$\tilde{\mu}$ determines occupied states via Aufbau filling.   Spectral projection can be carried 
out in a number of ways \cite{ANiklasson02A,ANiklasson03,RMcWeeny60,WClinton69,APalser98,GBeylkin99,KNemeth00,AHolas01}.
Of special interest here are recursive polynomial expansions of the projector, including the second 
order trace correcting (TC2) \cite{ANiklasson02A} and fourth order trace resetting (TRS4) 
\cite{ANiklasson03} purification algorithms.  These new methods (TC2 and TRS4) have convergence 
properties that depend only weakly on the band gap, do not require knowledge of the chemical potential 
and perform well for all occupation to state ratios. 
Perhaps most important to the current contribution, these methods converge rapidly to smooth, 
monotone projectors.

Prior to Ref.~[\onlinecite{COchsenfeld04}], Perturbed Projection demonstrated linear scaling in 
computation of the first electric polarizability for three-dimensional water clusters with the 
Hartree-Fock model \cite{VWeber04}. Also, in a preceding paper, we outlined a non-orthogonal density matrix
perturbation theory \cite{ANiklasson05a} for response to a change in basis (i.e.~as occurs in 
the evaluation of higher order geometric energy derivatives \cite{RAmos89}). 
In this article, the Perturbed Projection method is extended to higher orders
in the electric polarizability, up to fourth order in the total energy.  

This paper is organized as follows: First we describe the perturbation expansion and 
the computation of response properties through solution of the CPSCF equations.
Then we present extension of Perturbed Projection through higher orders and the 
computation of properties using a density matrix analogue of Wigner's $2 n+1$ rule.
Next,  we present several examples of calculated higher order response properties.
We show the saturation of hyperpolarizabilities up to  fourth order (i.e. up to the 
second hyperpolarizability $\gamma$) for a series water chains.
We also  demonstrate linear scaling complexity for the solution of the higher order CPSCF
equations and an approximate exponential decay in elements of higher order response functions
for 3D water clusters.  Finally, we discuss these results and present our conclusions. 

\section{The Coupled Perturbed Self-Consistent-Field Equations}

The Coupled-Perturbed Self-Consistent-Field (CPSCF) equations yield
static response functions and properties in models including both the 
Hartree-Fock (HF) and Density Functional Theory (DFT).  In the following
we develop Perturbed Projection for solution of the CPSCF equations in the framework of polarization and 
Hartree-Fock theory.  In many cases, the extension of Perturbed Projection to
the computation of other static perturbations is straightforward.  In
the case of model chemistries that involve DFT, an extra programming effort is required \cite{Lee_1994,PSalek02}.
Also, in the case of properties such as the NMR chemical shift and geometric derivatives 
(force constants), perturbation of the non-orthogonal basis requires additional 
considerations that we have detailed in a preceding paper \cite{ANiklasson05a}.

\subsection{Notation}

Superscripts and subscripts refer to perturbation order and 
self-consistent cycle count respectively. The symbols $\mathcal{D},\mathcal{F},\dots$
are matrices in an orthogonal representation, while
$D,F,\dots$ are the corresponding matrices in a non-orthogonal basis.
The transformation between orthogonal and non-orthogonal 
representations is carried out in ${\cal O}(N)$ using
congruence transformations \cite{JWilkinson65,GStewart73} provided 
by the approximate inverse (AINV) algorithm for computing  sparse 
approximate inverse Cholesky factors with a computational complexity
scaling linearly with the system size \cite{MBenzi95,MBenzi96,MBenzi01}.

\subsection{Response expansions}

Within HF theory, the total electronic energy $E_{\rm tot}$ of 
a molecule in a static electric field $\mathcal{E}$ is
\begin{equation}
  \begin{split}\label{totalE}
   E_{\rm tot}(\E)&=Tr \{ D \cdot (h^0+\mu\E) \}+\frac{1}{2}Tr \{ D \cdot (J[D]+K[D]) \} \\
                  &=Tr \{ D \cdot F[D] \}-\frac{1}{2}Tr \{D \cdot (J[D]+K[D]) \},
  \end{split}
\end{equation}
where $D \equiv D[\E]$ is the density matrix in the electric field $\mathcal{E}$, 
$h^0$ is the core Hamiltonian, $\mu$ is the dipole moment matrix, 
$J[D]$ is the Coulomb matrix, $K[D]$ the exact HF exchange matrix
and 
\begin{equation}
F \equiv F[\E]=h^0+\mu\E+J[D(\E)]+K[D(\E)]
\end{equation}
is the Fockian.
The total energy of a molecule in a homogeneous electric field may 
be developed in a Taylor expansion series around $\E = 0$ as
\begin{equation}
  \begin{split}
    E_{\rm tot}(\E)= E_{\rm tot}(0) 
    &-\sum_a\mu_a\E^a\\
    &-\frac{1}{2!}\sum_{ab}\alpha_{ab}\E^a\E^b\\
    &-\frac{1}{3!}\sum_{abc}\beta_{abc}\E^a\E^b\E^c\\
    &-\frac{1}{4!}\sum_{abcd}\gamma_{abcd}\E^a\E^b\E^c\E^d\\
    &+\dots,
  \end{split}
\end{equation}
 where $\alpha_{ab}$ is the polarizability, and $\beta_{abc}$ and 
 $\gamma_{abcd}$ are the first and second 
 hyperpolarizabilities, respectively, $\mu_a$ is the dipole 
 moment, and $\E^a$ is the electric field in direction $a$. 
 The polarizability $\alpha_{ab}$ is the second order response 
 of the total energy with respect to variation in the electric field 
 while the higher derivatives, $\beta_{abc}$ and $\gamma_{abcd}$, give 
 rise to the first and second hyperpolarizabilities \cite{HSekino86,SKarna91} where 
 \begin{subequations}\label{pol}
   \begin{gather}
     \alpha_{ab}=
     -\frac{\partial^2 E_{\rm tot}}{\partial \mathcal{E}^a\partial \mathcal{E}^b}
     \bigg|_{\mathcal{E}=0}=
     -2Tr[D^a\mu_b],\\
     \beta_{abc}=
     -\frac{\partial^3 E_{\rm tot}}{\partial \mathcal{E}^a\partial \mathcal{E}^b\partial \mathcal{E}^c}
     \bigg|_{\mathcal{E}=0}=
     -4Tr[D^{ab}\mu_c],\\
     \gamma_{abcd}=
     -\frac{\partial^4 E_{\rm tot}}{\partial\mathcal{E}^a\partial\mathcal{E}^b\partial \mathcal{E}^c\partial \mathcal{E}^d}
     \bigg|_{\mathcal{E}=0}=
     -12Tr[D^{abc}\mu_d].
   \end{gather}
\end{subequations}
Here $D^{a\ldots}$ denotes a density matrix derivative with respect to a field in directions $a\ldots$ 
at $\mathcal{E} = 0$.  The  density matrix derivative or ``response function'' is given by 
\begin{equation}
 \displaystyle\D^{a\ldots}=
 \frac{\partial^N}{\partial\E^{a\ldots}}\theta(\tilde{\mu}I-
 \F(\E))\bigg|_{\E=0} \label{DDeriv1}.
\end{equation}
 The Fockian may also be expanded order by order in the perturbation to yield
\begin{equation}\label{FockianTaylor}
  \begin{split}
    \F(\E)=\F^{0} & +\sum_a \F^{a}\E^{a}\\
    &+\frac{1}{2!}\sum_{ab} \F^{ab}\E^{a}\E^{b}\\
    &+\frac{1}{3!}\sum_{abc} \F^{abc}\E^{a}\E^{b}\E^{c}+\dots ~,
  \end{split}
\end{equation}
where $\F^{a}$ stands for $\partial\F(\E)/\partial\E^{a}$,
$\F^{ab}=\partial^2\F(\E)/\partial\E^{a}\partial\E^{b}$,
and so on for the higher order terms.
A similar expansion also holds for the density matrix $\D(\E)$.

Within HF theory, the unperturbed Fockian $F^0$ in the non-orthogonal basis is 
\begin{equation}
F^0=h^0+J(D^0)+K(D^0), \label{fockian0}
\end{equation}
while the first variation of the Fockian is 
\begin{equation}
F^a=\mu_a+J(D^a)+K(D^a)
\end{equation}
and the higher terms are given by 
\begin{equation}
F^{ab\ldots}=J(D^{ab\ldots})+K(D^{ab\ldots}). \label{fockianN}
\end{equation}
In computation of the unperturbed Fockian, the Coulomb matrix $J$ may be computed in ${\cal O}(N{\rm lg}N)$ 
with the Quantum Chemical Tree Code (QCTC) \cite{MChallacombe97} and the
exchange matrix $K$ computed in ${\cal O}(N)$ with the ${\cal O}(N)$-exchange  (ONX) algorithm 
that exploits quantum locality of the density matrix $D^0$ \cite{ESchwegler97}.
Likewise the Fockian derivatives, $F^{a\ldots}$, may be computed 
with the same algorithms in linear scaling time if elements of 
$D^{a\ldots}$ manifest an approximate exponential decay with atom-atom separation, 
similar to the decay properties of $D^0$. 

While the expansions above are given explicitly for Hartree-Fock Theory, similar expressions 
hold also for Kohn-Sham and hybrid HF/DFT,  which involve variation of the  exchange-correlation 
matrix $V_{xc}^{a\ldots}(D^0,D^a,\ldots)$ \cite{Lee_1994,PSalek02}.

\subsection{Conditions for self-consistency}\label{SelfConsistency}

The derivative density matrices and derivative Fockians depend on 
each other implicitly, and must be solved for self-consistently
via the CPSCF equations.
The necessary and sufficient criteria for convergence of the 
CPSCF equations involve generalized self-consistence conditions \cite{Furche_2001},
\begin{gather}
    [\F^{0} ,\D^{0}]=0,\label{eq:commutators1}\\
    [\F^{a} ,\D^{0}]+[\F^{0},\D^{a}]=0,\label{eq:commutators2}\\
  \begin{split}
    [\F^{ab},\D^{0}]&+\frac{1}{2}[\F^{a},\D^{b}]+\frac{1}{2}[\F^{b},\D^{a}] \\
    &+[\F^{0},\D^{ab}]=0,\label{eq:commutators3}\\
  \end{split}\\
  \begin{split}
    [\F^{abc},\D^{0}]&+\frac{1}{3}[\F^{ab},\D^{c}]+\frac{1}{3}[\F^{ac},\D^{b}]\\
    &+\frac{1}{3}[\F^{ab},\D^{c}]+\frac{1}{3}[\F^{a},\D^{bc}]\\
    &+\frac{1}{3}[\F^{b},\D^{ac}]+\frac{1}{3}[\F^{c},\D^{ab}]\\
    &+[\F^{0},\D^{abc}]=0,\label{eq:commutators4}\\
  \end{split}
\end{gather}
in addition to generalized idempotency-like constraints \cite{Furche_2001},
\begin{gather}
  \D^{0} =\D^{0} \D^{0},\label{eq:anticommutators1}\\
  \D^{a} =\{\D^{a},\D^{0}\},\label{eq:anticommutators2}\\
  \D^{ab}=\{\D^{ab},\D^{0}\}+\frac{1}{2}\{\D^{a},\D^{b}\}\label{eq:anticommutators3}\\
  \begin{split}
    \D^{abc}=\{\D^{abc},\D_0\}&+\frac{1}{3}\{\D^{ab},\D^{c}\}+\frac{1}{3}\{\D^{ac},\D^{b}\}\\
    &+\frac{1}{3}\{\D^{bc},\D^{a}\}\label{eq:anticommutators4},
  \end{split}
\end{gather}
where the anti-commutator notation $\{A,B\} = AB+BA$ has been used.

\section{Solving the higher order CPSCF equations with Perturbed Projection}

In solution of the CPSCF equations, it is first necessary to determine the ground state density matrix $\mathcal{D}^0$.  This may 
be accomplished in ${\cal O}(N)$ using a purification algorithm such as Niklasson's \cite{ANiklasson02A} 
second order trace correcting scheme (TC2) in conjunction with sparse atom-blocked linear algebra 
\cite{ANiklasson03,MChallacombe00B}.  Linear scaling is achieved for insulating systems through 
the dropping (filtering) of atom-atom blocks with Frobenious norm below a numerical threshold 
($\tau \sim 10^{-4}-10^{-7}$).  At SCF convergence the TC2 algorithm generates a polynomial sequence 
defining the ground state projector, from which the derivative density matrices are directly obtained.

Having solved the groundstate SCF equations, solution of the CPSCF equations commences with a guess at the 
derivative densities, followed by computation of derivative Fockians.  At the $r^{\rm th}$ 
CPSCF cycle, the $n^{\rm th}$ order derivative Fockians are 
\begin{equation}
    F^{a\ldots}_{r}= \left\{
    \begin{array}{ll}
      \mu_a+J(D^{a}_r)+K(D^{a}_r), & n=1\label{FockBuild}\\
      J(D^{a\ldots}_r)+K(D^{a\ldots}_r), & n>1 \,.
    \end{array}\right.
\end{equation}
After construction of the derivative Fockians, response functions through 
$D^{a\ldots}_{r+1}$ are computed, constituting one cycle in solution of the CPSCF.  
As described in Section \ref{ResponseFunctions},  these response functions are 
obtained directly through variation of the occupied subspace projector,  
\begin{equation}
    \displaystyle\D^{a\ldots}_{r+1}=
    \frac{\partial^n}{\partial\E^{a\ldots}}\theta(\tilde{\mu}I-
    \F_r(\E))\bigg|_{\E=0} \, , \label{DDeriv}
\end{equation}
which is accomplished via the Niklasson and Challacombe density matrix 
perturbation theory \cite{ANiklasson04}.  

After a few CPSCF cycles, the approach to self-consistency may be accelerated with 
Weber and Daul's DDIIS algorithm \cite{VWeber03}, 
\begin{equation}
    \displaystyle\widetilde{\F}^{a\ldots}_{r}=\sum_{k=r-s}^{r}c_k \F^{a\ldots}_{k} \label{DDIISEq} \\
\end{equation}
in which the $c_k$ coefficients are chosen to minimize the 
$n^{\rm th}$ order commutation 
relations, as in Eqs.~(\ref{eq:commutators2})-(\ref{eq:commutators4}). The application of the
DDIIS algorithm to acceleration of higher order CPSCF equations is developed further in Section \ref{DDIIS}.

At self-consistency, the conditions given in Section \ref{SelfConsistency} have been met, and it is then 
appropriate to compute response properties.   In general, we can use the expectation value 
\begin{equation}
E^{(\gamma)} = 2 (\gamma-1)! Tr(D^{(\gamma-1)} h^{(1)}), \label{Np1Rule}
\end{equation}
where in the case of polarization we have the (hyper)polarizabilities $\alpha_{ab}=-2\Tr[D^a\mu_b]$, 
$\beta_{abc}=-4\Tr[D^{ab}\mu_c]$ or $\gamma_{abcd}=-12\Tr[D^{abc}\mu_d]$.
Alternatively, it is possible to construct density matrix analogues of 
Wigner's $2 n+1$ rule, which allows the evaluation of response properties up to order $2 n+1$ from response 
functions of order $n$.  These analogues are given in Section \ref{Wigner2Np1} through $4^{th}$ order in
the energy.

\subsection{Perturbed Projection}\label{ResponseFunctions}

Although a number of analytic, asymptotically discontinuous representations exist for the Heaviside 
step function $\theta$, direct representation (and variation) of these forms as in Eq.~(\ref{DDeriv}) 
is problematic. Polynomial expansion of the step function is a alternative choice, but demands a very high 
order and can be costly.  Specifically, polynomial expansion of $\theta$ with a $p$'th order polynomial 
incurs a cost that is at best ${\cal O}(\sqrt{p})$ \cite{WLiang04}.  Polynomial expansion techniques,
such as those based on the the Tchebychev polynomials, may also be plagued by Gibbs oscillations \cite{AVoter96},
which are high order ripples in the approximate $\theta$ due to incompleteness.
Alternatively, recursive purification methods achieve high order representation in 
${\cal O}( \log p )$ \cite{ANiklasson03}.  Also, purification methods (such as TC2 and TRS4) 
yield projectors that are smooth and strictly monotonic,  

Each Perturbed Projection sequence is based on a corresponding  purification scheme or generator, such as TC2 \cite{ANiklasson02A}.   
The Perturbed Projection sequence is obtained by collecting terms of the response order by 
order upon perturbative expansion of its generator \cite{ANiklasson04}.
Perturbed Projection provides explicit, recursive formulae 
for the construction of response functions, retaining the convergence properties,  smoothness and 
montonicity of the generating sequence.   These explicit formulae stand in contrast to methods where the 
density matrix derivatives are implicitly defined as solutions to equations of Sylvester type \cite{Ochsenfeld97,HLarsen01a,COchsenfeld04}.

Sufficient to compute fourth order properties using the $2 n+1$ rule presented in Section \ref{Wigner2Np1}, 
Perturbed Projection is outlined in the following for computation of the second order response function.  
The Perturbed Projection sequence is started with the  $\X^{a\ldots}_{0}$, which are 
prepared from the Fockians $\F^0$, $\F^a$, and $\F^{ab}$ by  compressing their spectrum into the domain of 
convergence \cite{ANiklasson02A} using
\begin{equation}
    \X^0_{0}=\frac{\F_{max}-\F^0}{\F_{max}-\F_{min}} 
\end{equation}
and 
\begin{equation}
    \X^{a\ldots}_{0}=\frac{\F^{a\ldots}_{n}}{\F_{min}-\F_{max}},
\end{equation}
where $\F_{min}$ and $\F_{max}$ are upper and lower bounds to the eigenvalues of $\F^0$.  

While Perturbed Projection can be formulated within any purification scheme, we focus here on the
simple and efficient TC2 method \cite{ANiklasson02A}.  Briefly, TC2 
constructs a ground state projector through a series of trace correcting projections;  
when the trace is larger than $N_{\rm e}$, $x^2$ is used to reduce the trace, and 
when the trace is less than  $N_{\rm e}$, $2 x-x^2$ is used to increase the trace.  
The resulting sequence of correcting projections yields a step at the correct chemical potential. 
Within this framework, the second order TC2 Perturbed Projection sequence is 
\begin{equation}\label{PP1}
\left.
\begin{array}{ll}
\X^{ab}_{i+1}&=\{\X^{ab}_{i},\X^0_{i}\}+\frac{1}{2}\{\X^a_{i},\X^b_{i}\}\\
\X^a_{i+1}&=\{\X^a_{i},\X^0_{i}\}\\
\X^b_{i+1}&=\{\X^b_{i},\X^0_{i}\}\\
\X^0_{i+1}&=(\X^0_{i})^2 
\end{array} 
\right\}\,  {\rm Tr}[\mathcal{X}^0_{i}]\ge N_e 
\end{equation}
or 
\begin{equation}\label{PP2}
\left.
\begin{array}{ll}
      \X^{ab}_{i+1}&=2 \X^{ab}_{i}-(\{\X^{ab}_{i},\X^0_{i}\}+\frac{1}{2}\{\X^a_{i},\X^b_{i}\})\\
      \X^b_{i+1}&=2 \X^b_{i}-\{\X^b_{i},\X^0_{i}\} \\
      \X^a_{i+1}&=2 \X^a_{i}-\{\X^a_{i},\X^0_{i}\} \\
      \X^0_{i+1}&=2 \X^0_{i}-(\X^0_{i})^2
\end{array} 
\right\}\, {\rm Tr}[\mathcal{X}^0_{i}]< N_e.
\end{equation}
As with the  TC2 generator, the $n^{\rm th}$ order response functions
\begin{equation}
 \D^{a...} = n!\lim_{i\rightarrow\infty} \X_i^{a...},
\end{equation}
converge quadratically, reaching  convergence when either 
the error $\varepsilon = |\Tr[\X^0_{i}]-N_e| + |\Tr[\X^{a}_{i}]| + \ldots$, or 
the maximum element in the change $\delta \X^{a\ldots} = |\X^{a\ldots}_{i+1}-\X^{a\ldots}_{i}|$ 
falls below $\tau$, the atom-atom block drop tolerance described in Ref.~[\onlinecite{ANiklasson03}].
As described more completely in Ref.~[\onlinecite{ANiklasson05a}], when the solution gets 
close to convergence, i.e.~$|\Tr[\X^0_{i}]-N_e|< \epsilon$ with 
$\epsilon \approx 10^{-1}-10^{-3}$, we alternate the projection at each step,
which protects the convergence under the incomplete sparse linear algebra.

\subsection{Derivative DIIS}\label{DDIIS}

Direct inversion in the iterative subspace (DIIS), introduced
some time ago by Pulay \cite{Pulay80,Pulay82}, accelerates convergence toward 
self-consistency.   DIIS employs information accumulated during preceding 
iterations to construct an effective Fockian $\widetilde \F_{k}$ 
at the $k$-th SCF cycle, which minimizes the commutation error between the Fockian
and the density matrix. The effective Fockian is then used instead of $\F_{k}$
to generate an improved density matrix.  

Recently, Weber and Daul have developed the Derivative DIIS (DDIIS) scheme for accelerating 
convergence of the CPSCF equations \cite{VWeber03}.  Like DIIS, DDIIS is based on 
minimization of the Frobenious norm of an error matrix 
\begin{equation}
  \widetilde e_r^{a\ldots}=\sum_{i=r-s}^{r}c_i e_i^{a\ldots},
\end{equation}
where the $e_i^{a\ldots}$'s are just the $n$-th order commutator relation
of Eqs.~(\ref{eq:commutators1}-\ref{eq:commutators4}) (e.g. the first order error matrix 
is given by $e_i^{a}=[\F^{a}_i ,\D^{0}]+[\F^{0},\D^{a}_i]$).
The optimal coefficients $c_i$ are solutions to the quadratic programming problem
\begin{equation}
  \inf \left \{-\frac{1}{2}\sum_{i,j=r-s}^nc_iB_{ij}c_j,\, \sum_{i=r-s}^r c_i=1 \right \},
\end{equation}
where elements of the $\mathbf{B}$ matrix are given by 
$B_{ij}=\Tr[e_i^{a\ldots}(e_j^{a\ldots})^T]$.
A working equation is then obtained through the associated Euler-Lagrange equation
\begin{equation}\label{eq:diismatrix}
 \left ( \begin{array}{cc}
     \mathbf{B}     & \mathbf{1} \\
     \mathbf{1}^{T} & 0 
   \end{array}\right )
 \cdot \left ( \begin{array}{c}
     \mathbf{c}     \\
     \lambda  
   \end{array}\right )
  =  \left ( \begin{array}{c}
     \mathbf{0}      \\
         1  
   \end{array}\right ) \, ,
\end{equation}
 where $\mathbf{0}=(0,\ldots,0)^{T}$ and $\mathbf{1}=(1,\ldots,1)^{T}$ are
 vectors whose components are 0 and 1 respectively 
 and $\lambda$ is the Lagrange multiplier of the constraint 
 $\sum_{i=n-m}^{n}c_{i}=1$. The set of linear equations is solved
 by inverting the left-hand side matrix.  In the event of a singular or near singular
 matrix, the rank of Eq.~(\ref{eq:diismatrix}) is reduced by discarding the oldest entries (increasing $s$)
 until the linear system stabilizes.

\subsection{Density matrix formulation of Wigner's $2n+1$ rule }\label{Wigner2Np1}

Wigner's $2n+1$ rule, traditionally predicated on derivatives of the wavefunction,
yields order $2n+1$ in the energy response from $n$-th order derivatives \cite{SKarna91,SEpstein74}. 
A density matrix analogue of Wigner's $2n+1$ rule for a single perturbation parameter
was given to third order by McWeeny \cite{RMcWeeny62} and up to fourth order by 
Niklasson and Challacombe \cite{ANiklasson04} in the orthogonal representation.  
For completeness, we present non-orthogonal generalizations up to fourth order, 
which may require less memory due to the less dense structure of non-orthogonal matrix intermediates.

The first and second order energy corrections are well known, corresponding simply to expectation
values as in Eq.~(\ref{Np1Rule}).  Beyond second order, the $2 n+1$ rule offers a valuable alternative. 
The third order non-orthogonal contribution is 
\begin{align}
    E^{abc}
    &=2\sum_{P(a,b,c)}\Tr[[D^a,D^0]_SSD^bF^c]\label{eq:2n+1 third order}
\end{align}
where $P(a,b,c)$ stands for the permutation operator such that all
permutations of $a$, $b$ and $c$ are made (e.g. $P(a,b,c)$ generates the sum of
all the six terms: $(a,b,c)$, $(a,c,b)$, $(b,a,c)$, $(b,c,a)$, $(c,a,b)$ and $(c,b,a)$)
and $[A,B]_S=ASB-BSA$ where $S$ is the overlap matrix.  Similarly,  the fourth order non-orthogonal 
contribution is 
\begin{align}
  \begin{split}
    E^{abcd} =&\sum_{P(a,b,c,d)}\Tr[[D^{ab},D^0]_S SD^cF^d + \\ 
    &[D^a,D^0]_S S(D^{bc}F^d +D^bF^{cd})]\label{eq:2n+1 fourth order}
  \end{split}
\end{align}

For the orthogonal case $S=I$ and $D^a,F^c,\ldots$ are replaced by $\D^a,\F^c,\ldots$.
In most cases the complexity of these equations
can be reduced by taking advantage of indicial symmetry; 
$a,b,c,$ and $d$ represent the Cartesian directions $x,y,z$
so that terms with indecies in the same direction simplify. 
For example, $E^{aaaa}$ reduces to only one term requiring only 15 matrix multiplications.
In the worst case, where all the directions are different, i.e. $E^{aabc}$ 
(or any other permutation of $(a,a,b,c)$), the relation \eqref{eq:2n+1 fourth order} 
reduces to include only 12 terms with 180 matrix multiplications. 
Similar reductions of the computational cost also apply to 
Eq. \eqref{eq:2n+1 third order}. The number of matrix-matrix multiplies can be further reduced
if one uses an orthogonal representation, but this typically involves matrix-matrix multiplies 
with more dense intermediates.

\commentoutA{

\begin{table}[h]
  \centering
  \caption{\protect
    The longitudinal polarizability, $\alpha_{zz}$, for water chains at the RHF/6-31G level
    of theory, computed with {\sc MondoSCF} using {\tt GOOD} and {\tt TIGHT} numerical thresholds, 
   and also with the {\sc GAMESS} quantum chemistry package \cite{gamess}.
  }\label{tab:Alpha_1D_Values}
  \begin{tabular}{cccc}
    \toprule
    $N_{\rm H_2 O}$ &\multicolumn{1}{c}{{\sc GAMESS}}
        &\multicolumn{1}{c}{{\tt GOOD}}
        &\multicolumn{1}{c}{{\tt TIGHT}}\\
    \hline
    1 & 5.8136 & 5.813620 & 5.813588 \\
    2 & 6.3448 & 6.345037 & 6.344822 \\
    3 & 6.5844 & 6.584658 & 6.584435 \\
    4 & 6.7276 & 6.727905 & 6.727672 \\
    5 & 6.8226 & 6.823290 & 6.822857 \\
   10 & 7.0308 & 7.031056 & 7.030858 \\
   15 & 7.1047 & 7.104904 & 7.104770 \\
   20 & 7.1424 & 7.142580 & 7.142422 \\
    \botrule
  \end{tabular}
\end{table}

\begin{table}[h]
  \centering
  \caption{\protect
    The longitudinal first hyperpolarizability, $\beta_{zzz}$,
    for water chains at the RHF/6-31G level of theory, computed with 
    {\sc MondoSCF} using {\tt GOOD} and {\tt TIGHT} numerical thresholds, 
    and also with the {\sc GAMESS} quantum chemistry package \cite{gamess}.
  }\label{tab:Beta_1D_Values}
  \begin{tabular}{cccccc}
    \toprule
    $N_{\rm H_2 O}$ &\multicolumn{1}{c}{{\sc GAMESS}}
    &\multicolumn{1}{c}{{\tt GOOD}}
    &\multicolumn{1}{c}{{\tt GOOD}\footnote[1]{The density matrix based 2n+1 rule has been used.}}
    &\multicolumn{1}{c}{{\tt TIGHT}}
    &\multicolumn{1}{c}{{\tt TIGHT}$^a$} \\
    \hline
     1 & -30.6125 & -30.611029 & -30.612627 & -30.612163 & -30.612256 \\
     2 & -29.5444 & -29.547427 & -29.548504 & -29.544907 & -29.544994 \\
     3 & -25.3696 & -25.372208 & -25.373615 & -25.370297 & -25.370381 \\
     4 & -22.1411 & -22.143436 & -22.145040 & -22.141494 & -22.141603 \\
     5 & -19.8925 & -19.902088 & -19.904449 & -19.896462 & -19.897141 \\
    10 & -14.8063 & -14.807075 & -29.617990 & -14.806973 & -14.807119 \\
    15 & -12.9713 & -12.969238 & -12.972227 & -12.971940 & -12.972124 \\
    20 & -12.0334 & -12.028709 & -12.033633 & -12.034014 & -12.034238 \\
    \botrule
  \end{tabular}
\end{table}

\begin{table}[h]
  \centering
  \caption{\protect
    The longitudinal second hyperpolarizability, $\gamma_{zzzz}$,
    for water chains at the RHF/6-31G level of theory, computed with 
    {\sc MondoSCF} using {\tt GOOD} and {\tt TIGHT} numerical thresholds, 
    and also with the {\sc GAMESS} quantum chemistry package \cite{gamess}.
  }\label{tab:Gamma_1D_Values}
  \begin{tabular}{rccccc}
    \toprule
    $N_{\rm H_2 O}$ &\multicolumn{1}{c}{{\sc GAMESS}}
    &\multicolumn{1}{c}{{\tt GOOD}}
    &\multicolumn{1}{c}{{\tt GOOD}\footnote[1]{The density matrix based 2n+1 rule has been used.}}
    &\multicolumn{1}{c}{{\tt TIGHT}}
    &\multicolumn{1}{c}{{\tt TIGHT}$^a$} \\
    \hline
     1 &  330.5753 & 330.54375 & 330.54319 & 330.57193 &  330.5724 \\
     2 &  820.1398 & 820.19231 & 820.19667 & 820.14775 &  820.1493 \\
     3 & 1008.5656 & 1008.5855 & 1008.6073 & 1008.5752 & 1008.5765 \\
     4 & 1103.4813 & 1103.5053 & 1103.5280 & 1103.4883 & 1103.4902 \\
     5 & 1168.9563 & 1169.2104 & 1169.2531 & 1169.0630 & 1169.0754 \\
    10 & 1324.2906 & 1325.1208 & 1324.6321 & 1324.2975 & 1324.2999 \\
    15 & 1381.8657 & 1383.0802 & 1382.2322 & 1381.8758 & 1381.8738 \\
    20 & 1411.4264 & 1414.4092 & 1411.8528 & 1411.4410 & 1411.4292 \\
    \botrule
  \end{tabular}
\end{table}

\begin{figure}[h]
  \caption{\protect
    Total CPU time of the fifth CPSCF iteration of fourth order for
    the water cluster sequence with the 6-31G and 6-31G** 
    basis sets and the {\tt GOOD} and {\tt TIGHT} 
    numerical thresholds (see text) controlling numerical
    precision of the result. The lines are fits to the 
    last three and four points, respectively.
  }\label{fig:Gamma_scaling}
  \resizebox*{3.6in}{!}{\includegraphics[angle=-90.00]{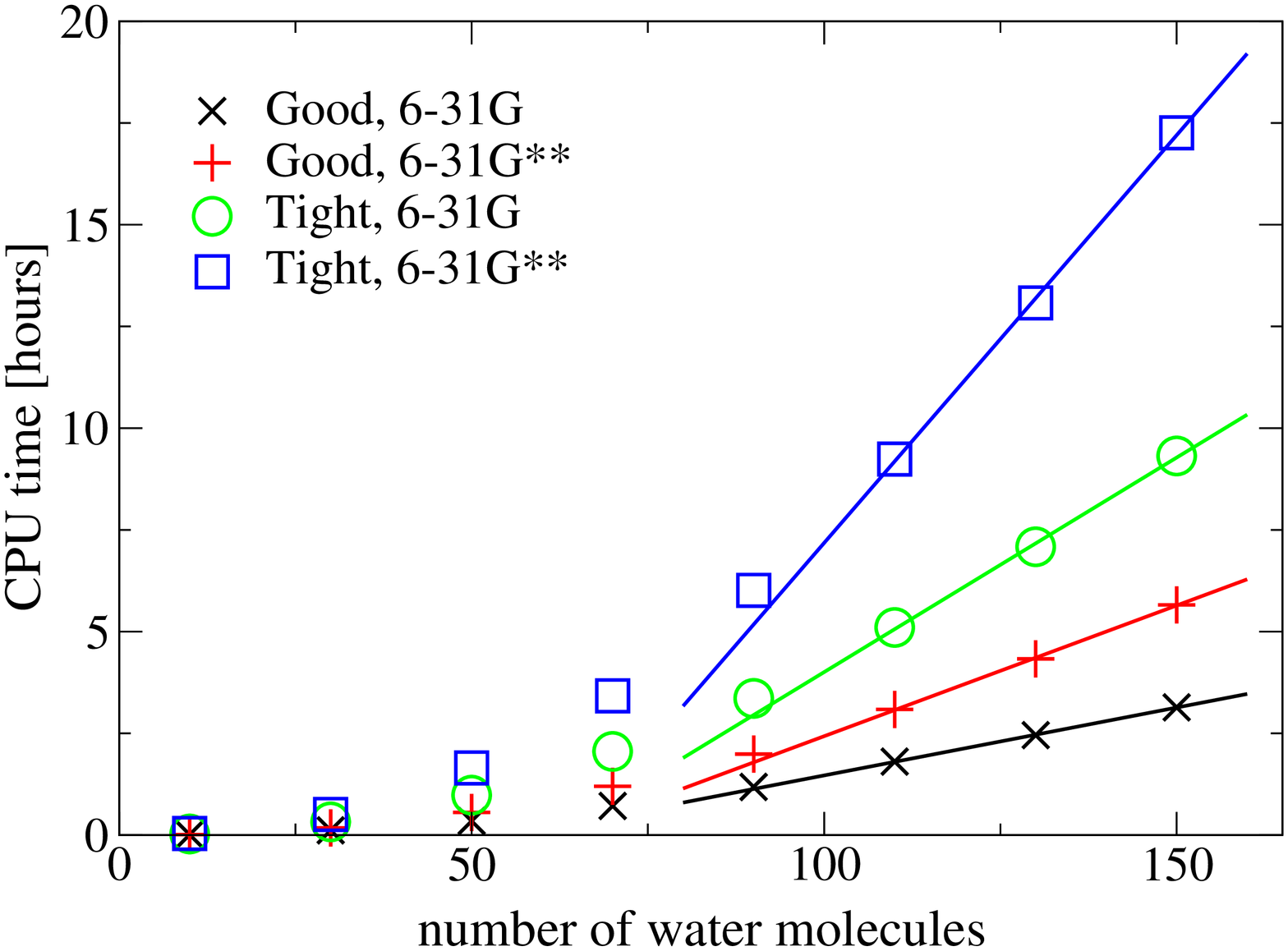}}
\end{figure}

\begin{figure}[h]
  \caption{\protect
    QCTC CPU time of the fifth CPSCF iteration of fourth order for
    the water cluster sequence with the 6-31G and 6-31G** 
    basis sets and the {\tt GOOD} and {\tt TIGHT} 
    numerical thresholds (see text) controlling numerical
    precision of the result. The lines are fits to the 
    last three and four points, respectively.
  }\label{fig:Gamma_QCTC_Timing}
  \resizebox*{3.6in}{!}{\includegraphics[angle=-90.00]{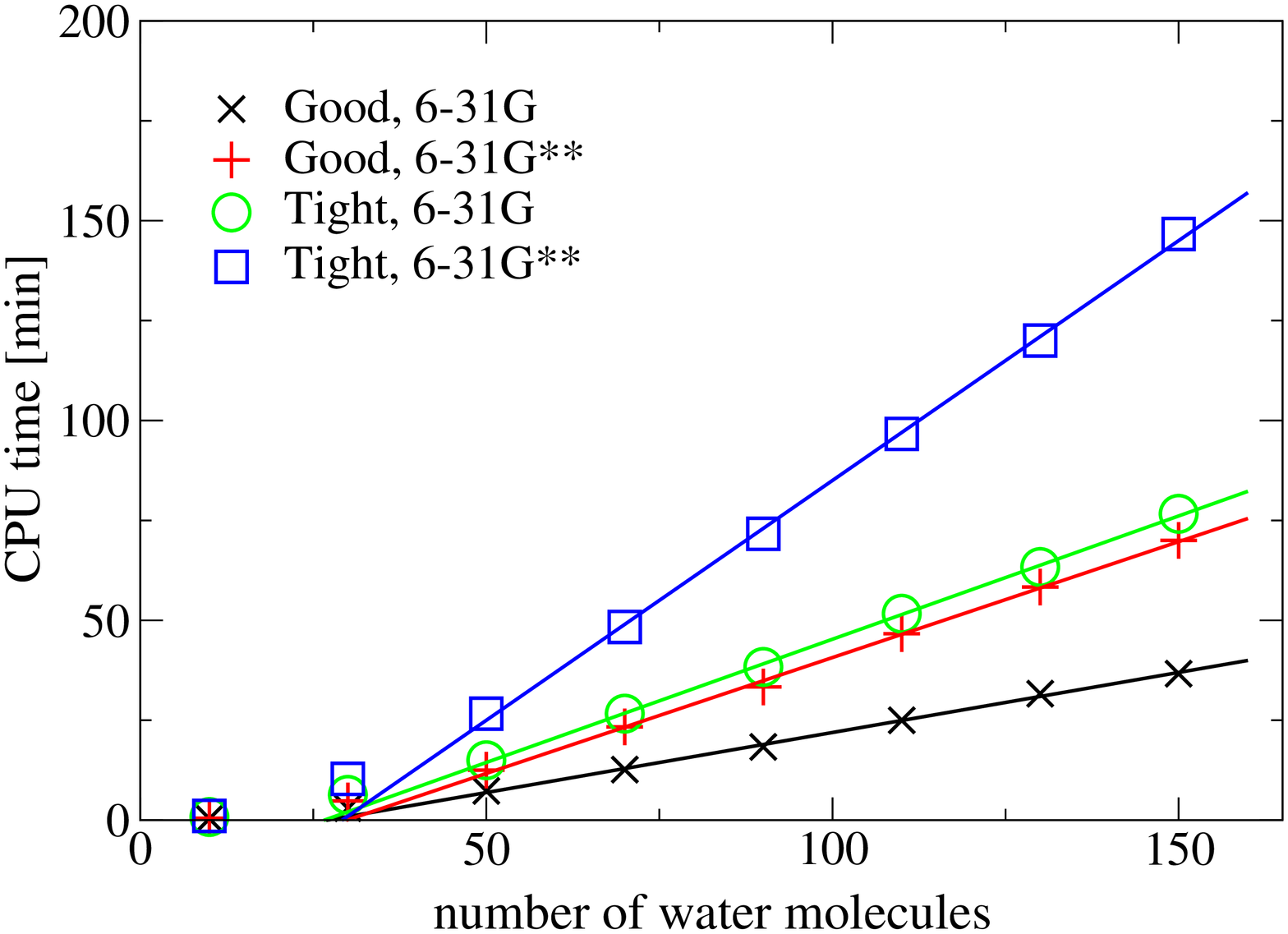}}
\end{figure}

\begin{figure}[h]
  \caption{\protect
    TC2 CPU time of the fifth CPSCF iteration of fourth order for
    the water cluster sequence with the 6-31G and 6-31G** 
    basis sets and the {\tt GOOD} and {\tt TIGHT} 
    numerical thresholds (see text) controlling numerical
    precision of the result. The lines are fits to the 
    last three and four points, respectively.
  }\label{fig:Gamma_TC2R_Timing}
  \resizebox*{3.6in}{!}{\includegraphics[angle=-90.00]{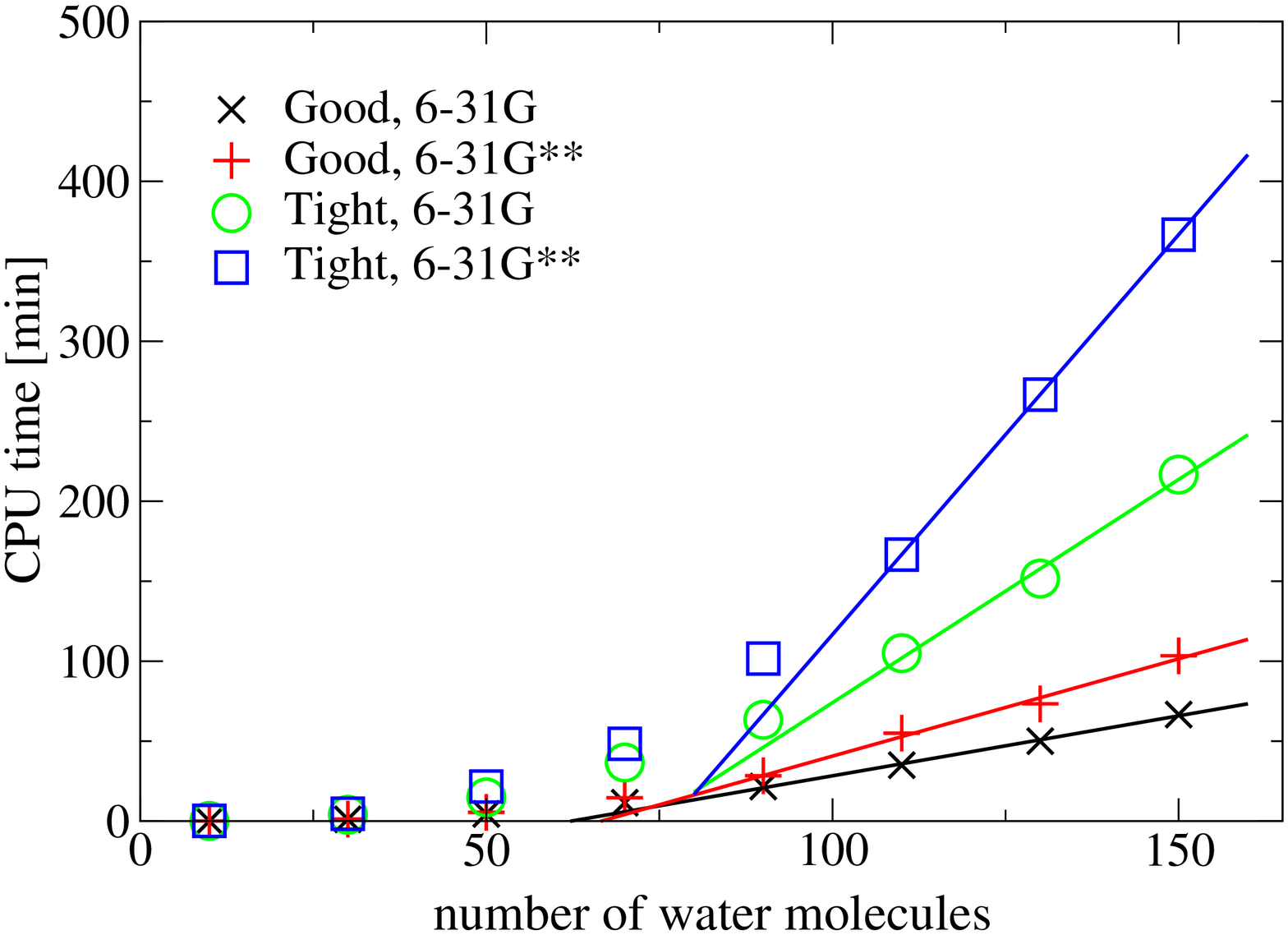}}
\end{figure}

\begin{figure}[h]
  \caption{\protect
    ONX CPU time of the fifth CPSCF iteration of fourth order for
    the water cluster sequence with the 6-31G and 6-31G** 
    basis sets and the {\tt GOOD} and {\tt TIGHT} 
    numerical thresholds (see text) controlling numerical
    precision of the result. The lines are fits to the 
    last three and four points, respectively.
  }\label{fig:Gamma_ONX_Timing}
  \resizebox*{3.6in}{!}{\includegraphics[angle=-90.00]{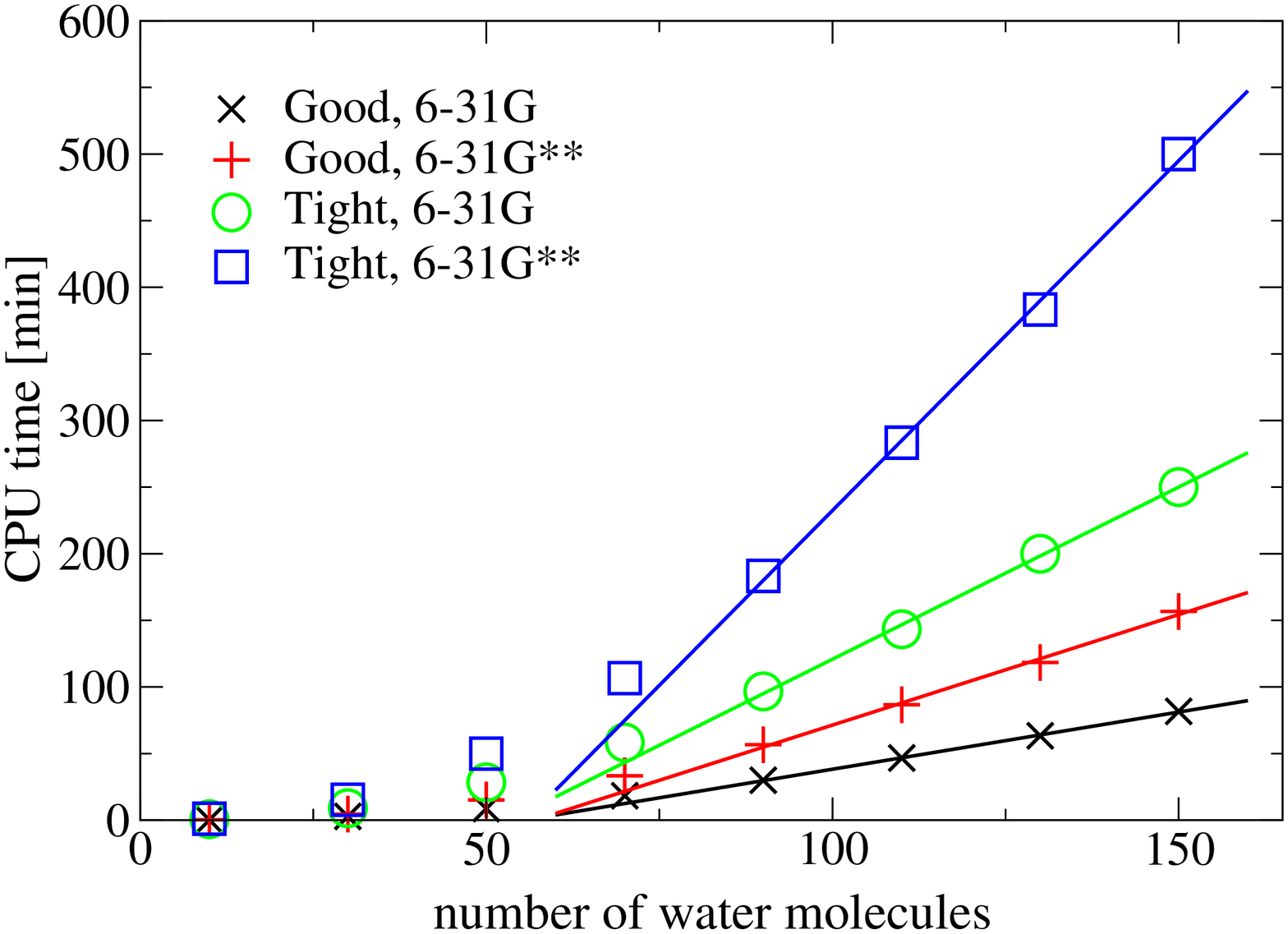}}
\end{figure}

\begin{figure}[h]
  \caption{Total CPU times with increasing order of the response for 
           the fifth CPSCF cycle computed as the $n+1$ expectation value, Eq.~(\ref{Np1Rule}).}\label{TimeWithOrder}
  \resizebox*{3.6in}{!}{\includegraphics[angle=-90.00]{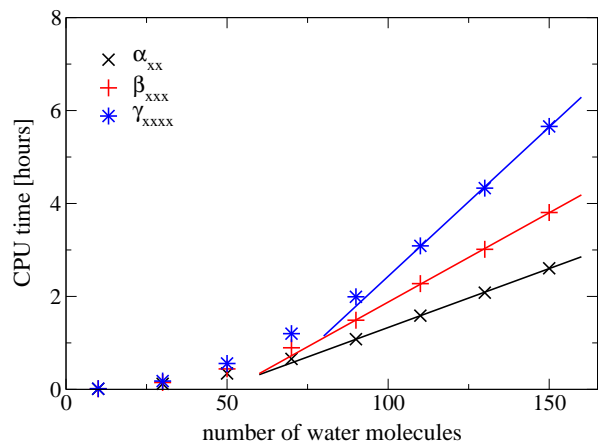}}
\end{figure}

\begin{figure}[t]
  \caption{\protect
    Superposition of the magnitudes of the RHF/6-31G density matrix
    derivative elements $D_{cd}$, $D^{x}_{cd}$, $D^{xx}_{cd}$ and $D^{xxx}_{cd}$
    along the $x$ axis with the separation of basis function centers
    for $\rm (H_2O)_{150}$. The density matrix 
    derivatives have been converged to within {\tt TIGHT} (e.g. 
    a matrix threshold $\tau=10^{-6}$ $[a.u.]$).
  }\label{fig:Superposition_Decay}
  \resizebox*{3.6in}{!}{\includegraphics[angle=-90.00]{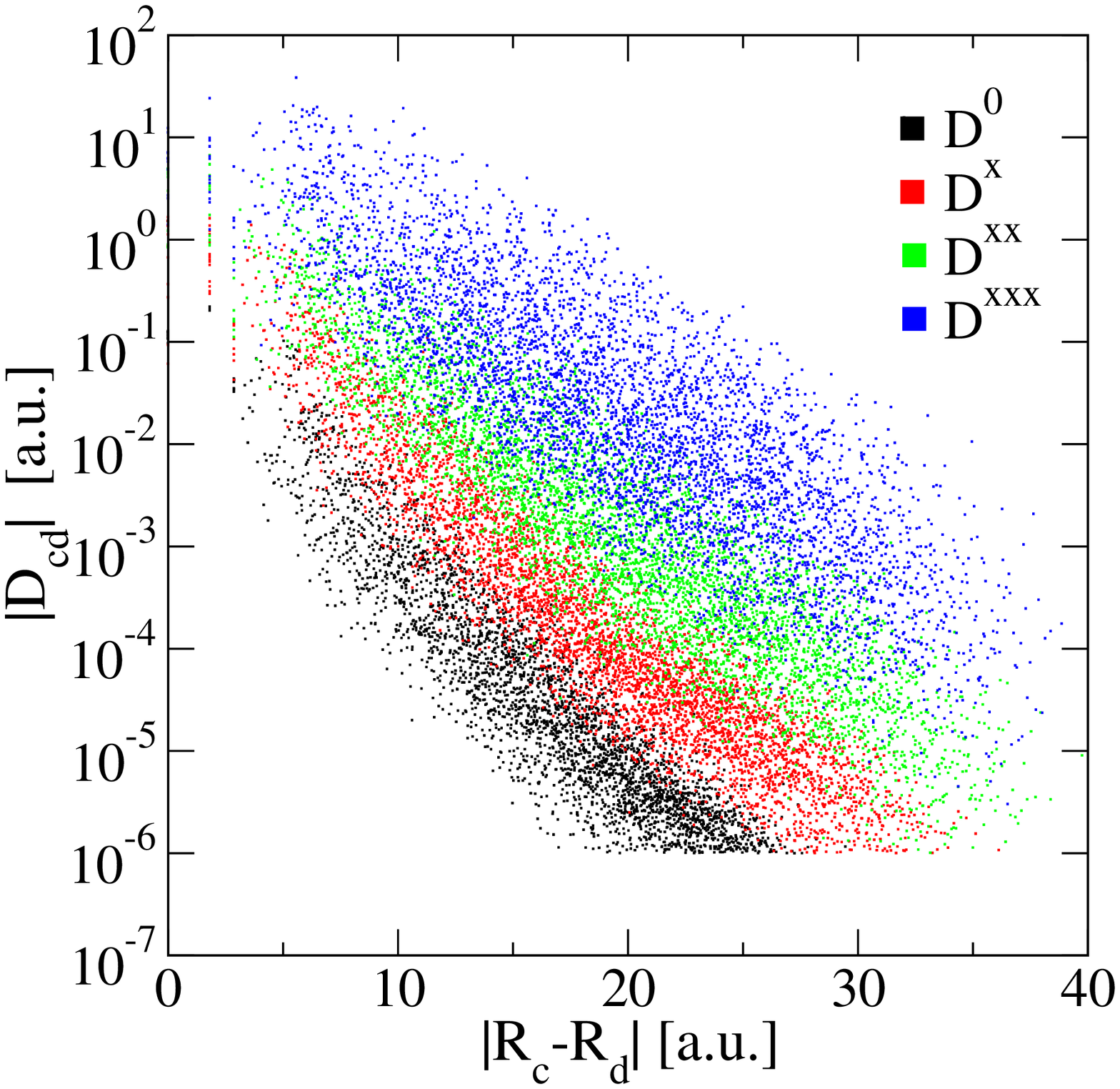}}
\end{figure}

}

\section{Results} \label{RESULTS}

We have implemented these methods in the MondoSCF suite of linear scaling quantum chemistry 
programs \cite{MondoSCF}.  The construction of the Fockian and derivative Fockian was carried
out using the linear scaling QCTC method for computation of the Coulomb matrix \cite{MChallacombe97,CTymczak04a} and the
ONX algorithm \cite{ESchwegler97,CTymczak04b} for computation of the Hartree-Fock exchange matrix. 
The CPSCF equations were solved in an entirely orthogonal representation,
with higher order properties evaluated using the $2 n+1$ rule with the non-orthogonal formulae given
by Eqs.~(\ref{eq:2n+1 third order}-\ref{eq:2n+1 fourth order}).  Two different levels of numerical 
accuracy have been used, {\tt GOOD} and {\tt TIGHT}.   Thresholds that define the {\tt GOOD} 
accuracy level include a matrix threshold $\tau=10^{-5}$, as well as other numerical thresholds 
detailed in Ref.~[\onlinecite{CTymczak04a}], which deliver 6 digits of relative accuracy in the total 
energy.  The {\tt TIGHT} option involves the matrix threshold $\tau=10^{-6}$ and delivers 8 
digits of relative accuracy in the total energy.  

Calculations were carried out on a single Intel Xeon 2.4GHz processor running RedHat Linux 8.0 and  
executables compiled with Portland Group Fortran Compiler pgf90 4.0-2 \cite{PGF90}.

Convergence of the CPSCF equations for the water systems described in the following 
are typically achieved in about 10 cycles, independent of cluster size, basis set,
matrix threshold or order of the response calculated.

All results are reported in atomic units.  Also, unless otherwise noted, all timings and values 
have been obtained by computing the $n^{\rm th}$ order response function and evaluation with the
$n+1$ rule (expectation value), Eq.~(\ref{Np1Rule}).
\subsection{One dimensional water chains}

Perturbed Projection has been used to compute the (hyper)polarizabilities $\alpha_{zz}$, 
$\beta_{zzz}$ and $\gamma_{zzzz}$ of linear water chains up to (H$_2$O)$_{20}$. 
These calculations have been carried out with {\sc MondoSCF} at the RHF/6-31G level of theory using 
both the {\tt GOOD} and {\tt TIGHT} thresholding parameters,  as well as with the conventional algorithms implemented
in the {\sc GAMESS} quantum chemistry package \cite{gamess}.  These static properties have 
been evaluated at the geometries given by Otto {\em et al.}~\cite{POtto99}, and the GAMESS
results are given to the number of digits provided by that program.
The {\sc MondoSCF} results have been obtained both as expectation values, given by Eq.~(\ref{Np1Rule}), 
and using the non-orthogonal density matrix $2 n+1$ rules given in 
Eqs.~(\ref{eq:2n+1 third order}-\ref{eq:2n+1 fourth order}).

As a benchmark, we have also carried out calculations with the linear chain (H$_2$O)$_{20}$ with 
the {\tt VERYTIGHT} numerical thresholding parameters, which employ a $10^{-7}$ drop tolerance and
aim to provide 10 digits of precision in the total energy.   These {\tt VERYTIGHT} calculations yield  
$\alpha_{zz}=7.142422\,a.u.$, $\beta_{zzz}=-12.033362\,a.u.$ and $\gamma_{zzzz}=1411.425500\,a.u.$. 

\subsection{Linear scaling: 3D water clusters}

Linear scaling computation of the RHF/6-31G and RHF/6-31G** second hyperpolarizability,
achieved with Perturbed Projection, is shown for three-dimensional water clusters 
in Fig~\ref{fig:Gamma_scaling}.  These timings are the total CPU time for the fifth CPSCF cycle, 
including build time for $\F^{abc}$ ({\sc ONX} and {\sc QCTC}), 
iterative construction of $\D^{abc}$ (Perturbed Projection via {\sc TC2}) and all intermediate
steps including the congruence transformation.
A breakdown of the dominant contributions to these totals are given
in Figs~\ref{fig:Gamma_QCTC_Timing}-\ref{fig:Gamma_ONX_Timing}, which show timings for Coulomb 
summation ({\sc QCTC}), Perturbed Projection ({\sc TC2}), and exact exchange ({\sc ONX}).

Figure \ref{TimeWithOrder} shows the increase in cost associated with computing  higher order response functions.
Corresponding to this increase in cost, Fig.~\ref{fig:Superposition_Decay} shows the magnitude of atom-atom blocks 
of density matrix derivatives up to third order as a function of atom-atom distance 
when perturbed by a static electric field.  The density response shows an approximate 
exponential decay as a function of internuclear distance with the rate of  decay
slowing slightly and the distribution shifted up with increasing order in the perturbation.

\section{Discussion}

In our current formulation, the increase in magnitude and reduction of locality in elements of the 
response function makes achieving linear scaling more difficult with increasing order in perturbation. 
Nevertheless, linear scaling has been achieved at the HF level of theory up to fourth order (i.e.~$\gamma$) 
in the total energy for three-dimensional systems and  non-trivial basis sets.  At fourth order,
Perturbed Projection and exact exchange were the dominant costs in solving the CPSCF equations,
as shown in Figs~\ref{fig:Gamma_TC2R_Timing} and \ref{fig:Gamma_ONX_Timing}.  For the fourth order Perturbed Projection, 
$N$-scaling is achieved 
between 70 to 110 water molecules, depending on $\tau$.  Despite a nearly dense $D^{abc}$, the 
dominant work in its construction always involves multiplication with matrices that are significantly 
more sparse, as $\X^{abc}_i\X^{0}_i$ or $\X^{ab}_i\X^{c}_i$.   Likewise, $N$-scaling is achieved between 
70 to 90 water molecules for construction of the Hartree-Fock exchange contribution.  In this case, 
the approximate decay of the density matrix still leads to linear scaling through ordered skip out lists,
as described in Ref.~[\onlinecite{ESchwegler97}].  In both cases, the increase in response function magnitude
is equivalent to tightening numerical thresholds, which increases the cost and delays the onset of linear
scaling. 

In tables \ref{tab:Alpha_1D_Values}-\ref{tab:Gamma_1D_Values} we find that 
a reduction of the drop tolerance by one order of magnitude leads to an increase in precision by
1-2 significant digits, with {\tt GOOD} and {\tt TIGHT} yielding approximatively 3-4 and 5-6 
correct digits  independent of the response order.  We further observe about 1 extra digit of accuracy 
when using the $2n+1$ rule.  This might be expected from the higher order error propagation resulting from 
products of lower order response functions, relative to evaluation with  Eq.~(\ref{Np1Rule}), 
which involves an error that is always linear in a higher order derivative density matrix.  

As shown in Fig.~\ref{TimeWithOrder}, computing the second order response is significantly cheaper 
than the third order response, and involves an earlier onset of linear scaling. Because evaluation 
of properties with the $2 n+1$ rule is of negligible cost relative to solving the CPSCF equations, the cost 
difference for evaluating $\gamma$ with the $2 n+1$ rule relative to the $n+1$ expectation is just the 
difference (roughly 2:3) between the computation of $\beta$ and $\gamma$ shown in Fig.~\ref{TimeWithOrder}.  

\section{Conclusions}

Linear scaling has been demonstrated for the computation of response properties beyond second
order in the total energy using Perturbed Projection for solution of the Coupled-Perturbed Self-Consistent-Field 
equations.
In addition, we have provided details of 
the computational method, used three-dimensional systems and non-trivial basis sets
to demonstrate linear scaling and provided a (preliminary) assessment of 
error control.  Perturbed Projection for the 
computation of higher order response functions is quadratically convergent, simple 
to implement through higher orders and numerically stable.  Perturbed Projection is 
not unique to the Hartree-Fock model, the TC2 generator or the {\sc MondoSCF} 
$N$-scaling algorithms, but can be straightforwardly extended to models that include 
exchange correlation (DFT), other other purification schemes such as TRS4 \cite{ANiklasson03}
as well as other electronic structure programs.  

We have shown that response functions (density matrix derivatives) through fourth 
order are local upon {\em global} electric perturbation, corresponding to an 
approximate exponential decay of matrix elements.  However, the magnitude of the 
corresponding response functions increases with increasing perturbation order, equivalent 
to tightening of the matrix drop tolerance, $\tau$.   While we have not attempted to 
work out a detailed analysis for the propagation of error, it may be possible 
to develop a more effective thresholding scheme for high orders. In addition 
to being somewhat more accurate, the $2 n+1$ rule also provides a significantly 
cheaper alternative to the computation of expectation values and an earlier onset 
of linear scaling. 

A similar exponential decay in the first order response corresponding to a {\em local} nuclear 
displacement has likewise been demonstrated by Ochsenfeld and Head-Gordon \cite{Ochsenfeld97}. 
This behavior is expected to hold generally for both local and global perturbations to 
insulating systems.  Thus, the potential exists for Perturbed Projection to achieve linear scaling
for a large class of static molecular properties within the HF, DFT and hybrid HF/DFT model 
chemistries. Of particular interest, the recently developed non-orthogonal density matrix perturbation 
theory put forward in a proceeding article \cite{ANiklasson05a} may enable linear scaling computation of analytic 
second derivatives, which are important in computation of the Hessian matrix.  

\begin{acknowledgments}
 This work has been supported by the US Department of Energy 
 under contract W-7405-ENG-36 and the ASCI project.  
 The Advanced Computing Laboratory of Los 
 Alamos National Laboratory is acknowledged.
 All the numerical computations have been
 performed on computing resources located at this facility.
\end{acknowledgments}

\bibliography{Response3}

\end{document}